# DOTIFS: Fore-optics and calibration unit design

Haeun Chung*[a,b], A. N. Ramaprakash[c], Pravin Khodade[c], Chaitanya V. Rajarshi[c],
Sabyasachi Chattopadhyay[c], Pravin A. Chordia[c], Amitesh Omar[d], and Changbom Park[b]
[a]Seoul National University, Seoul, Korea
[b]Korea Institute for Advanced Study, Seoul, Korea
[c]Inter-University Centre for Astronomy and Astrophysics, Pune, India
[d]Aryabhatta Research Institute of Observational Sciences, Nainital, India

## ABSTRACT

We present fore-optics and calibration unit design of Devasthal Optical Telescope Integral Field Spectrograph (DOTIFS). DOTIFS fore-optics is designed to modify the focal ratio of the light and to match its plate scale to the physical size of Integral Field Units (IFUs). The fore-optics also delivers a telecentric beam to the IFUs on the telescope focal plane. There is a calibration unit part of which is combined with the fore-optics to have a light and compact system. We use Xenon-arc lamp as a continuum source and Krypton/Mercury-Neon lamps as wavelength calibration sources. Fore-optics and calibration unit shares two optical lenses to maintain compactness of the overall subsystem. Here we present optical and opto-mechanical design of the calibration unit and fore-optics as well as calibration scheme of DOTIFS.

## 1. INTRODUCTION

Devasthal Optical Telescope Integral Field Spectrograph (DOTIFS)[1] is a new multi-Integral Field Unit (multi-IFU) spectrograph for 3.6m Devasthal Optical Telescope[2]. It is designed and fabricated by instrumentation group at Inter-University Centre for Astronomy and Astrophysics (IUCAA) in Pune, India. The telescope is managed by Aryabhatta Research Institute of Observational Sciences (ARIES) at Nainital, India. Korea Institute for Advanced Study and Seoul National University are participating in this project as international collaborators. DOTIFS is designed to obtain spectra from the spatially resolved area on the sky. A combination of optical spectrographs and lenslet+fiber concept IFUs can obtain 2,304 spectra simultaneously from 370 to 740 nm range with spectral resolution 1200-2400. 16 IFUs are located at the side port of Cassegrain focus of the telescope and deployed on 8' focal plane by a novel IFU deployment system. Each IFU has 8.7 arcsec x 7.4 arcsec field of view, with an array of 12 by 12 hexagonal aperture microlenses and fibers behind. Details of the overall instrument can be found in the overview paper[1].

The instrument requires the fore-optics system to deliver appropriate light into IFUs so they can be transferred to the spectrograph optics properly. F-ratio of the light from the telescope primary and secondary mirrors is not proper to be accepted by IFUs. Thus a set of optics are required to modify the light appropriately. DOTIFS also requires a designated calibration system since the telescope does not have a calibration system to be used by DOTIFS spectrograph. By having a dedicated calibration unit, we can choose proper light source which is optimized to calibrate the DOTIFS spectrograph. Figure 1 shows an opto-mechanical overview of DOTIFS side port subsystem. Fore-optics and calibration unit are combined as an integrated module. It is attached directly to the Cassegrain side port entrance.

In this paper, we present optical and opto-mechanical design of fore-optics and calibration unit. First, we describe design and performance of the fore-optics subsystem. Second, we show details of calibration unit subsystem. We list detail of calibration light sources as well as calibration unit optical design. Third, we present opto-mechanical design of fore-optics and calibration unit, with a component which both parts share. Fourth, we describe calibration scheme of DOTIFS. Finally, we will end up with reporting the current status of the fore-optics and calibration unit followed by a summary of the paper.

*hchung@astro.snu.ac.kr; phone +82-10-7542-2737

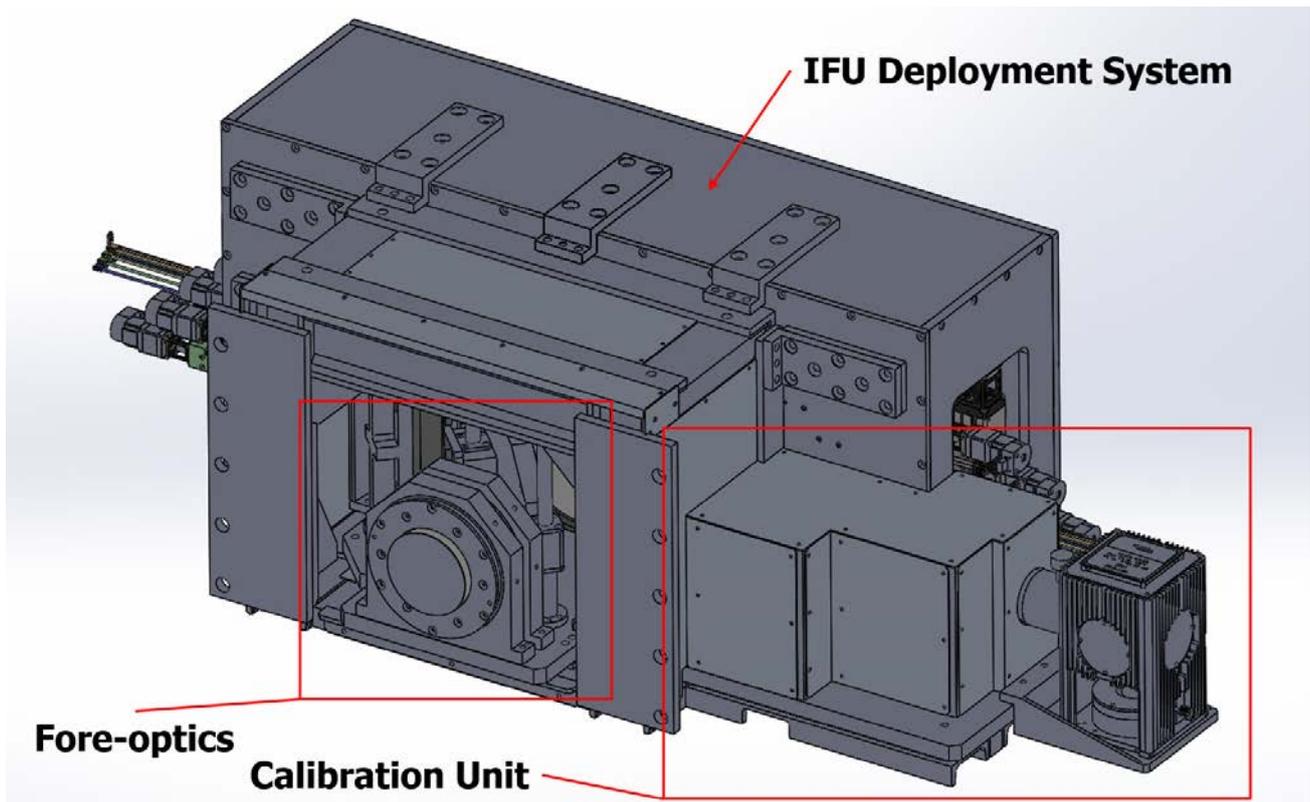

Figure 1 Overall mechanical rendering of DOTIFS subsystems, positioned at the telescope Cassegrain side port. Each subsystem is pointed out with red arrows.

## 2. FORE-OPTICS

### 2.1 Design goals

We design DOTIFS fore-optics to match plate scale on the focal plane to the microlens sampling size. DOTIFS requires a fore-optics to match IFUs with the original telescope plate scale, which is 157 μm/arcsec. To work without fore-optics, the physical size of one IFU would be only 1.37 mm x 1.16 mm, and the size of each microlens would be 125.6 μm. Considering available diameter of the optical fiber, packing fibers with 125.6 μm spacing is physically not feasible. Thus, we need to modify the plate scale on the focal plane and increase the physical size of the microlens accordingly. Our choice is using fore-optics to modify that plate scale on the focal plane to 300 μm/0.8 arcsec. This corresponds to the focal ratio of 21.486.

The optical performance of the fore-optics magnifier should be good enough to make image quality captured by IFU to be limited by sampling size. Root-Mean-Square (RMS) spot radius size less than 50 μm is estimated to satisfy the optic performance requirement. More than 80% of the energy from a 2 arcsec size source should be contained within the 750μm hexagonal area at the microlens array surface, which corresponds to 2.5 times of spatial sampling size. The optics should have a field of view as wide as possible while maintaining homogeneous imaging quality over the entire focal plane. Also, the output beam should be telecentric so incidence angle of the light to each IFU would not change over the optics field of view.

### 2.2 Optical design

We design fore-optics to refract telescope F/9 beam to F/21.486 telecentric beam. It changes plate scale on the focal plane from 157 μm / 1 arcsec to 300 μm / 0.8 arcsec. This is about 2.39 times of magnification. Original telescope field of view is known as 10 arcmins at the Cassegrain side port, but we utilize 8 arcmins of them. We use the optical layout of the 3.6m DOT to start fore-optics design and put several plane elements near the original focal plane as a starting design. We set merit function as a sum of RMS spot radius of three different wavelengths and five field points. Also, effective focal length is constrained to match the desired plate scale. We add several operands to make telecentric beam at the focus. We constrain the lens positions to be around the position of the original focal plane since we can put optics only around the Cassegrain side port space. During the optimization, we manually add or subtract a lens element to find the optimal solution.

The design (Figure 2) is comprised of five singlets, all spherical lenses. Three relatively small singlets are located close to the side port entrance, and two large singlets are located just before the focal plane. We have two Calcium Fluoride component to maximize the throughput at the near-UV (370 – 400 nm) wavelength range.

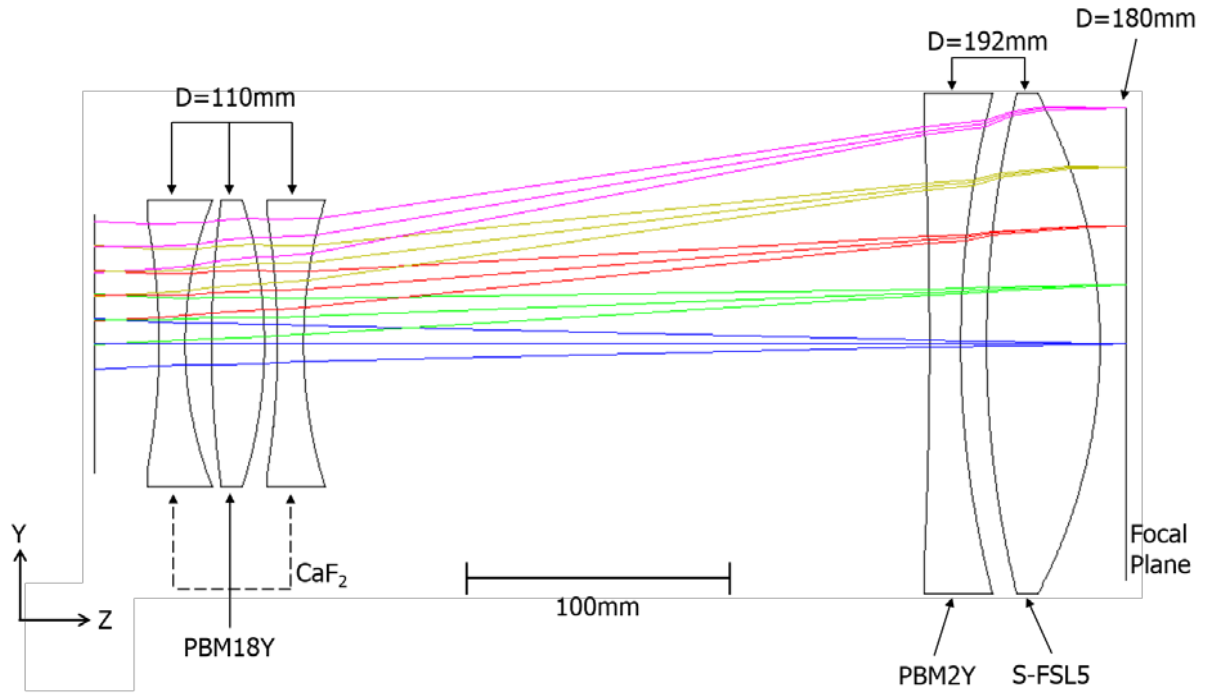

Figure 2 Fore-optics layout. Diameter and glass material of each component is shown. Different color lines represent light from five different field points. (0, 1, 2, 3, and 4 arcminutes)

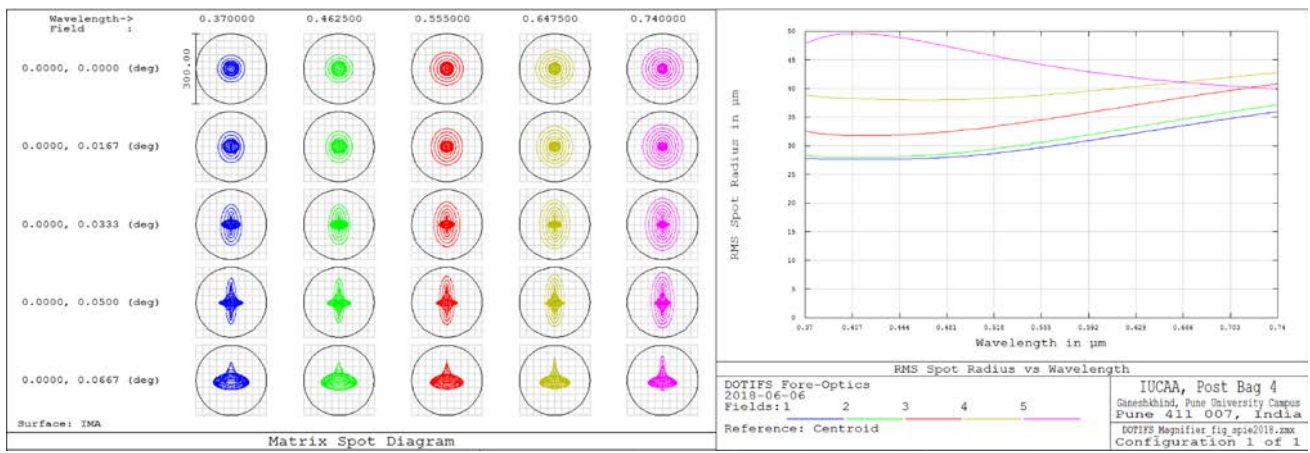

Figure 3 (Left) Matrix spot diagram of the fore-optics. Color represents different wavelength. (Right) Wavelength versus RMS spot radius plot. Color represents the light of different field points, as in the top of the bottom rows of the left figure.

## 2.3 Performance

In Figure 3, we show spot diagram and wavelength versus RMS spot radius graph of the fore-optics. The diameter of a black circle around each spot is 300 μm, which represent a size of one microlens. An actual microlens is a hexagonal shape with 300 μm vertex to vertex, which is slightly smaller than the black circle in the diagram. In general, the spots are well shaped and concentrated at the center, but their size is increasing as the location of field point gets further from the center. We find that this change is inevitable by doing magnification at the very end of converging light. There could be some effect on IFU imaging quality near the edge of the field, which needs to be mitigated by post-processing. There is also chromatic dispersion between the center and the edge of the 8' diameter field, which requires proper calibration and post-processing. Since the IFU field of view is less than 10 arcsec, there will be negligible chromatic dispersion effect within a single IFU. We plot extended source encircled energy fraction depend on wavelength at different field points in Figure 4. The result shows the fractions are mostly higher than 80%. We also check sensitivity analysis of fabrication and alignment tolerances on fore-optics, as well as thermal analysis. Since the optics has no moving parts, we use telescope secondary mirror as a compensator during the analysis. Due to the slow focal ratio of the optics, optical performance variation is negligible (< 5% change) during the analysis. Therefore we conclude the design satisfies the goal.

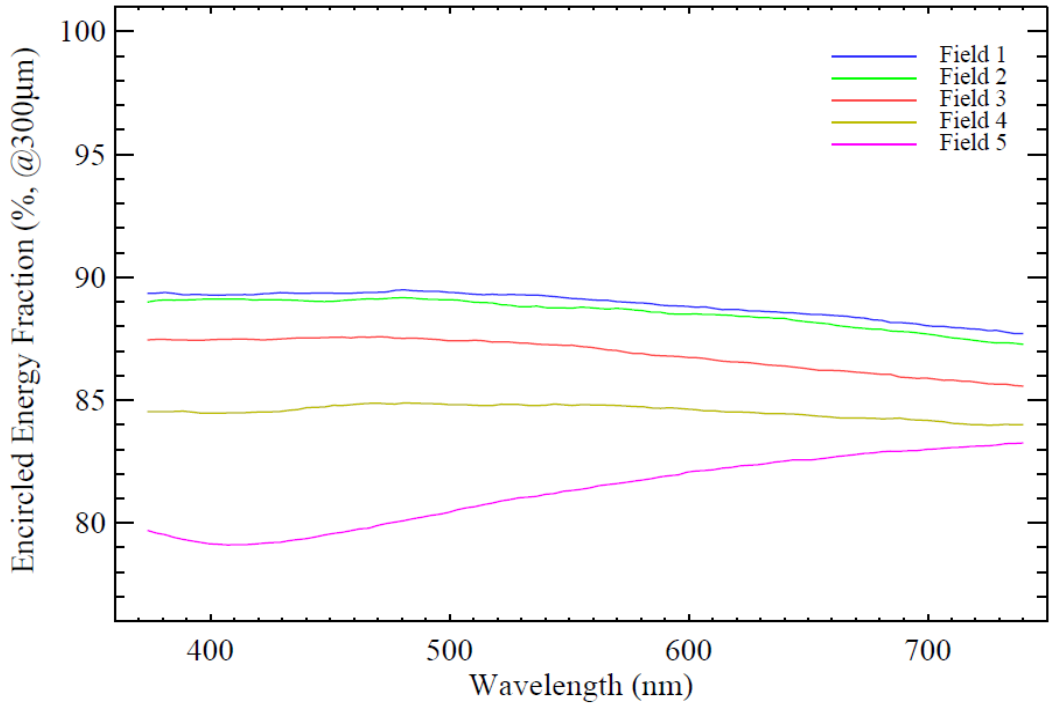

Figure 4 Extended source encircled energy fraction of the fore-optics with 2 arcsec diameter size uniform source. Sampling size is 750 μm in a circle, which correspond to 2.5 times of the microlens sampling size. Color represents energy fraction from different field points, as in the right side of Figure 3.

## 3. CALIBRATION UNIT

### 3.1 Design Goals

We design DOTIFS calibration unit to provide a telescope-alike beam to the focal plane from calibration light sources. CCD image of Integral Field Spectrograph is complex. Thus a precise calibration process to obtain high-quality data is required. Major calibration sources are continuum source and emission line source. Continuum source is used to map the position of the spectrum from each spatial elements (fibers), and emission line source is used to map the pixel location in a spectral direction to the wavelength. Geometrical property of the light from those calibration sources should be identical to the one from telescope and fore-optics, to have no systematic difference in between.

### 3.2 Calibration sources and integrating sphere

DOTIFS uses Xenon arc lamp (Newport part #6263) as a continuum source. We choose Xenon arc lamp because of its relatively flat radiation curve within the DOTIFS working wavelength range, compare to the traditional continuum sources such as Quartz lamp or Halogen lamp. It is expected to provide sufficient photons in the near UV range as well. Also, calibration with a continuum source can be done in seconds since the lamp has sufficient power. Krypton (Kr) and Mercury-Neon (HgNe) lamp (Newport part #6031 and #6034) are chosen as wavelength calibration sources, considering there emission line profiles and DOTIFS working wavelength range. The original supplier of all three lamps is Oriel Instrument which is a well-known lamp manufacturer in the field of astronomical instrumentation. We also employ a small integrating sphere (Newport part #819D-SL-2) to scramble the light from calibration source and provide uniform and collimated light to the calibration optics.

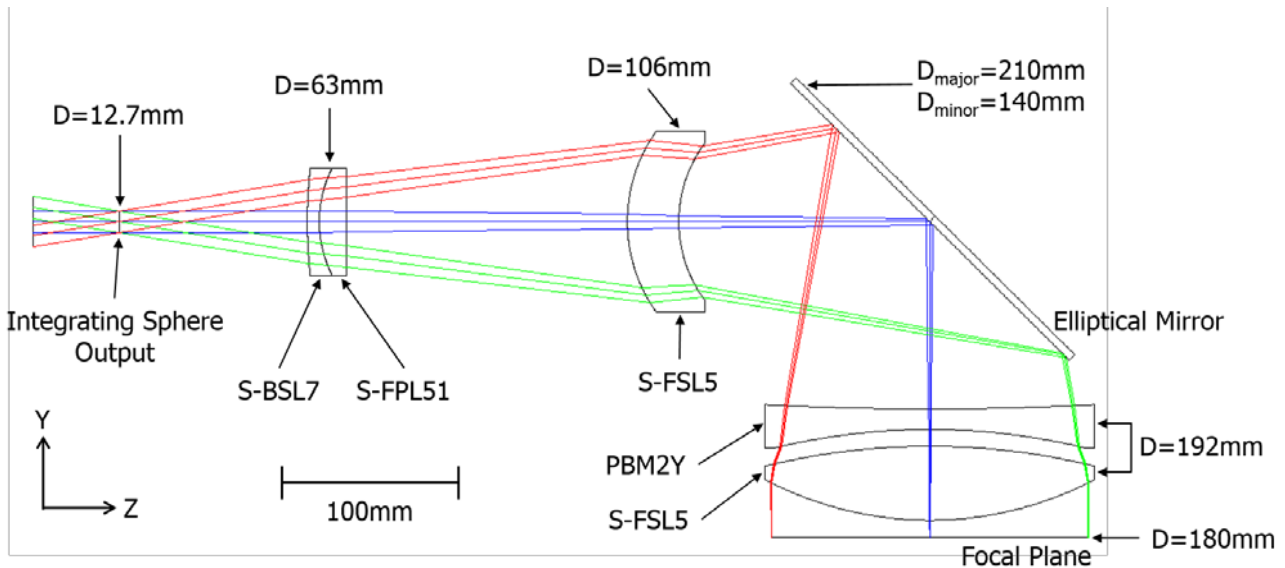

Figure 5 Layout of calibration unit optics. Elements from the integrating sphere output to the focal plane is shown. Diameter and glass material of each component are labeled with black arrows. The last two lenses in this diagram are identical to the last two lenses in Figure 2. Different color lines represent light from different angles.

### 3.3 Optical design

Light from calibration sources enters to the integrating sphere. The output light from the sphere is considered as a uniform light source and used as an object of calibration optics. Considering limited space around the focal plane which IFUs are deployed, calibration optics and fore-optics are designed to use two lenses in common, which are components just before the focal plane (Figure 5). We utilize available space of the Cassegrain side-port area and the gap among the fore-optics elements to locate remaining calibration optics. The movable large elliptical mirror is placed between the third and fourth lenses of the fore-optics lens and acting as a calibration source selector. One doublet and one singlet are placed next to the integrating sphere output as a dedicated calibration optics. The combination of optical components produces F/21.486 light at the fore-optics focal plane.

## 4. FORE-OPTICS AND CALIBRATION UNIT OPTO-MECHANICS

We present opto-mechanical structure of the fore-optics and calibration unit in Figure 6 and Figure 7. The design has two moving components, a round shape rotating mirror which selects the type of calibration light source and a movable elliptical mirror which can direct light from calibration sources to the focal plane. Rotating mirror and elliptical mirror are positioned by rotation stage and guideway linear stage, both manufactured by Holmarc Opto-Mechatronics. The rotating mirror is positioned in the middle of continuum source, emission line source, and the integrating sphere. It can select the light source by rotating 90°. The elliptical mirror can move in the direction perpendicular to the opto-mechanics base plate. Light from calibration source falls on focal plane when elliptical mirror blocks the light from the telescope side. There are baffles to separate the space before and after the integrating sphere physically. Thus, light can be transferred to the remaining optics only through the integrating sphere. An electronic shutter is placed at the input port of the integrating sphere so calibration sources can be turned on and warmed up while target observation is ongoing.

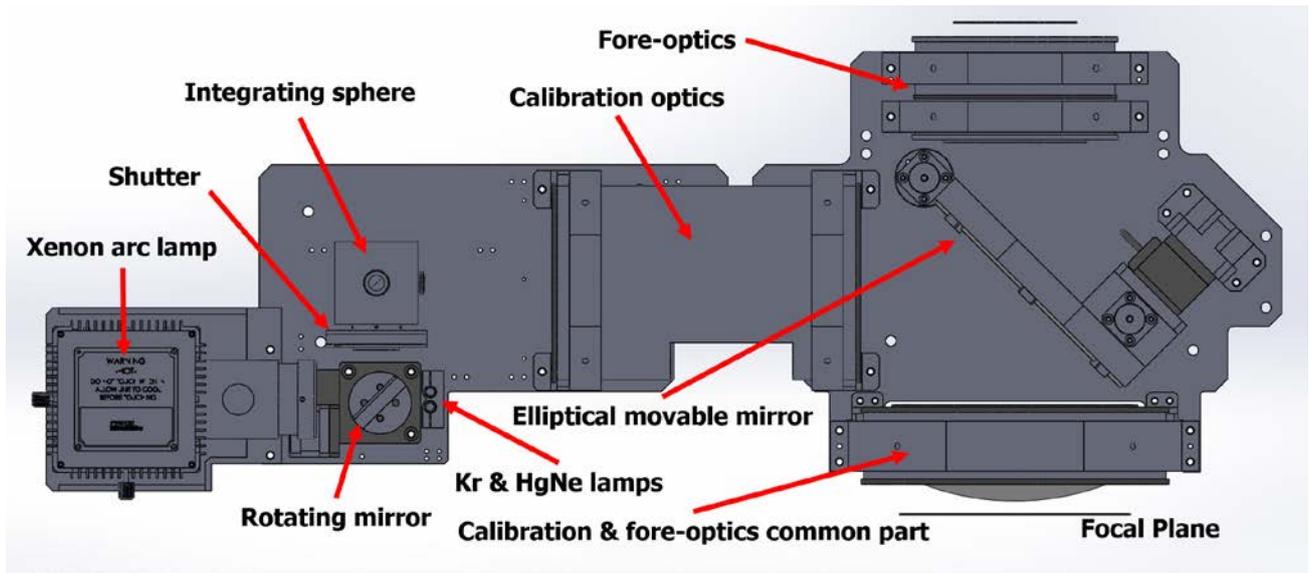

Figure 6 Top-view of DOTIFS fore-optics and calibration unit opto-mechanical structure. Location of subcomponents is pointed out with red arrows. Covers are removed to show internal components. The elliptical movable mirror can be retracted to the out of paper direction by linear guideway stage.

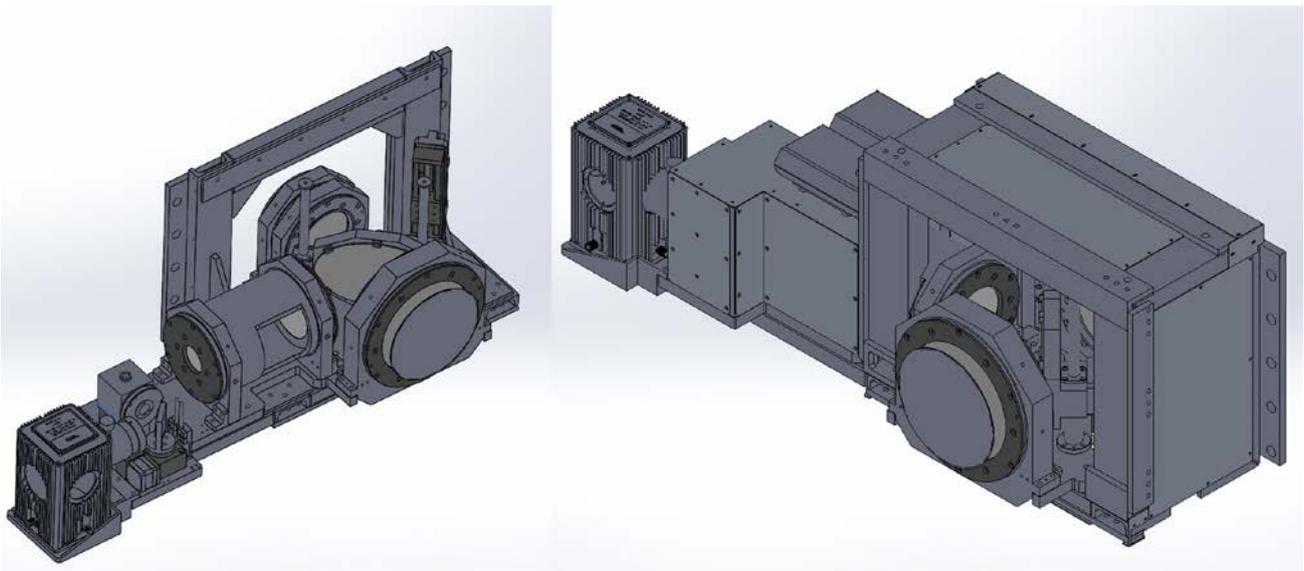

Figure 7 (Left) View of fore-optics and calibration unit at a different viewing angle. The 3-dimensional structure of the components can be recognized from this view. (Right) Fore-optics and calibration unit opto-mechanics with cover.

## 5. CALIBRATION SCHEME

Here we describe fundamental calibration scheme of DOTIFS. In principle, each science exposure requires two sets of calibration exposure, before and after the target observation. This is to have highest calibration accuracy as well as to trace possible flexure effect during the science exposure. One set of calibration exposure is comprised of one flat field image (with continuum source) and one wavelength calibration image. (With wavelength calibration sources). We make simulated calibration exposure images from DOTIFS data simulator[3], using known throughput of the instrument and radiation curve of continuum and emission line sources. The images are shown in Figure 8. At each calibration exposure, Xenon lamp will be turned on for few minutes ahead to be stabilized. A shutter near the integrating sphere input port will be closed during this warm-up phase. When Xenon lamp becomes stabilized, the shutter will be opened, and the elliptical mirror will be placed between the fore-optics lenses to block the light from the telescope. Continuum source image will be taken when everything is ready. After that, the arc lamp will be turned off, and wavelength calibration lamps will be turned on to take wavelength calibration source image. After taking two calibration images, the shutter will be closed, and the elliptical mirror is moved out so the next science exposure can be made.

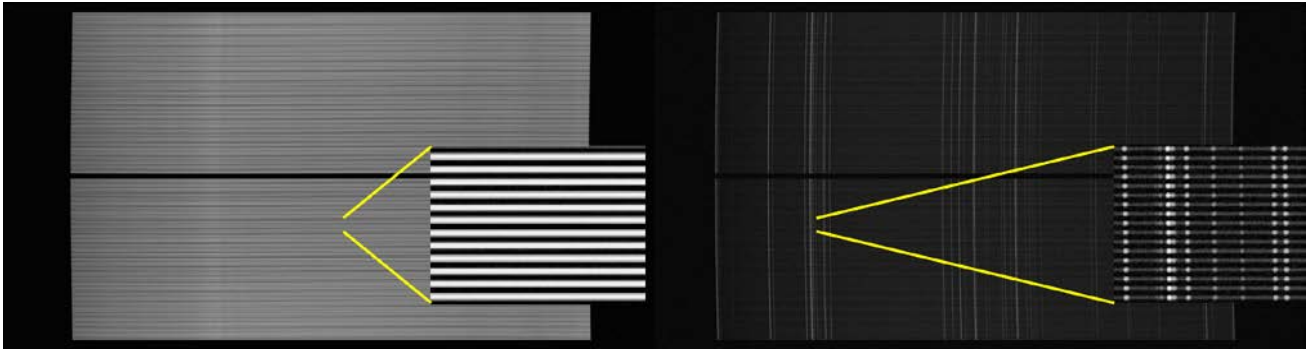

Figure 8 Simulated DOTIFS calibration image with (Left) Xenon arc lamp and (Right) Krypton and Mercury-Neon lamp. We also show the enlarged view of middle parts.

## 6. CURRENT STATUS

We describe current status regarding each subcomponent DOTIFS fore-optic and calibration unit as of 2018 May. We ordered lens components to the to the lens fabricator (Phoenix Optical Technologies, UK). We expect to receive them by the end of July 2018. We ordered and received round mirror (Off the shelf, Edmund optics) and elliptical mirrors (custom, Orion optics). We also purchased calibration light sources and integrating sphere (Newport). Rotation stage and linear guideway (Holmarc) also arrive at the laboratory. Opto-mechanical components are fabricated by local machines shops and ready to be assembled. Currently, we are working on verification of parts which are on our hands. We will start assembly of components once lens components arrive.

## 7. SUMMARY

We present the optical and opto-mechanical design of the DOTIFS fore-optics and calibration unit system. They are designed to share optical components to be accommodated by limited space around the telescope side port. Detailed design and structure are described. The system is expected to provide magnified beam and calibration light to the focal plane and IFUs. Currently, we ordered and received most of the components except optical lenses. We will start assembly of the system once we have lenses.